\documentclass[12pt]{iopart}
\usepackage{epsfig}
\usepackage{citesort}

\newcommand{\gtrsim}{\,\rlap{\lower3.7pt\hbox{$\mathchar\sim$}}
\raise1pt\hbox{$>$}\,}
\newcommand{\lesssim}{\,\rlap{\lower3.7pt\hbox{$\mathchar\sim$}}
\raise1pt\hbox{$<$}\,}

\begin{document}

\title{New cosmological mass limit on thermal relic axions}
\author{Steen Hannestad,$^{ 1}$ Alessandro Mirizzi,$^{ 2,3}$ 
and Georg Raffelt$^{ 2}$}
\address{$^1$ Department of Physics and Astronomy,
University of Aarhus,
Ny Munkegade, DK-8000 Aarhus C, Denmark\\}
\address{$^2$ Max-Planck-Institut f\"ur Physik
(Werner-Heisenberg-Institut)\\
F\"ohringer Ring 6, 80805 M\"unchen, Germany\\}
\address{$^3$ Dipartimento di Fisica
                and Sezione INFN di Bari\\
                Via Amendola 173, 70126 Bari, Italy\\}
\date{7 April 2005}

\begin{abstract}
  Observations of the cosmological large-scale structure provide
  well-established neutrino mass limits. We extend this argument to
  thermal relic axions. We calculate the axion thermal freeze-out
  temperature and thus their cosmological abundance on the basis of
  their interaction with pions.  For hadronic axions we find a new
  mass limit $m_a<1.05$~eV (95\%~CL), corresponding to a limit on the
  axion decay constant of $f_a>5.7\times 10^6$~GeV.  For other models
  this constraint is significantly weakened only if the axion-pion
  coupling is strongly suppressed. For comparison we note that the
  same approach leads to $\sum m_\nu<0.65$~eV (95\%~CL) for neutrinos.
\end{abstract}

\maketitle

\section{Introduction} 

Low-mass cosmic relic particles provide a hot-dark matter component of
the universe. After they decouple in the early universe, they freely
stream and decrease the spectral power of primordial density
fluctuations on small scales. This well-studied effect has been
established as a standard method to derive neutrino mass limits where
the results of different authors differ primarily in the used
observational data sets and assumed priors on some of the cosmological
parameters (for a recent review see Ref.~\cite{Hannestad:2004nb}). Two
of us have recently extended this method to several generic cases of
other low-mass thermal relics that may have been produced in the early
universe~\cite{Hannestad:2003ye}, for example weakly interacting
bosons such as the hypothetical axions.

For a generic low-mass particle species $X$ that was once in thermal
equilibrium, the crucial parameters that enter the structure-formation
argument are the assumed particle mass $m_X$, the number $g_X$ of
internal degrees of freedom, the relevant quantum statistics (boson
vs.\ fermion), and finally the number $g_{*}$ of effective cosmic
thermal degrees of freedom at the epoch when the $X$ particles
thermally decouple. These parameters imply the present-day cosmic mass
density $\Omega_X$ in these particles as well as their velocity
distribution and primordial free-streaming scale.

In Ref.~\cite{Hannestad:2003ye} two of us considered thermal relic
axions as a specific example and found an upper mass limit of
2--3~eV. The main purpose of the present paper is to sharpen this
result in two ways.  First, we use an updated set of observational
data. Second, we take a closer look at the axion freeze-out in the
early universe to obtain a more accurate value for $g_{*}$ in terms
of the appropriate coupling constants.
 
We begin in Sec.~\ref{sec:axions} with a brief summary of the
relevant aspects of axion physics and continue in
Sec.~\ref{sec:freeze-out} with a determination of the freeze-out
conditions as a function of the relevant axion coupling constants.  In
Sec.~\ref{sec:likelihood} we perform a likelihood analysis based on
recent cosmological data to determine the allowed range of axion
parameters. We finish in Sec.~\ref{sec:summary} with a summary and
interpretation of our results.

\section{Axions}                                    \label{sec:axions}

Quantum chromodynamics is a CP-violating theory, implying that the
neutron should have a large electric dipole moment, in conflict with
the opposite experimental evidence. The most elegant solution of this
``strong CP problem'' was proposed by Peccei and Quinn (PQ) who showed
that CP conservation is dynamically restored in the presence of a new
global U(1) symmetry that is spontaneously broken at some large energy
scale~\cite{Peccei:1977hh,Peccei:1977ur}.
Weinberg~\cite{Weinberg:1977ma} and Wilczek~\cite{Wilczek:1977pj}
realized that an inevitable consequence of the PQ mechanism is the
existence of a new pseudoscalar boson, the axion, which is the
Nambu-Goldstone boson of the PQ symmetry. This symmetry is explicitly
broken at low energies by instanton effects so that the axion acquires
a small mass. Unless there are non-QCD contributions, perhaps from
Planck-scale physics~\cite{Kamionkowski:1992mf,Barr:1992qq}, the mass
is
\begin{equation}\label{eq:axmass}
m_a=\frac{z^{1/2}}{1+z}\,\frac{f_\pi m_\pi}{f_a}
=\frac{6.0~{\rm eV}}{f_a/10^6~{\rm GeV}}\,,
\end{equation}
where the energy scale $f_a$ is the axion decay constant or PQ scale
that governs all axion properties, $f_\pi=93$~MeV is the pion decay constant,
 and $z=m_u/m_d$ is the mass ratio of
the up and down quarks.  We will follow the previous axion literature
and assume a value $z=0.56$ \cite{Gasser:1982ap,Leutwyler:1996qg}, but
we note that it could vary in the range 0.3--0.7
\cite{Eidelman:2004wy}.

The PQ scale is constrained by various experiments and astrophysical
arguments that involve processes where axions interact with photons,
electrons, and hadrons. The interaction strength with these particles
scales as $f_a^{-1}$, apart from model-dependent numerical factors.
If axions indeed exist, experimental and astrophysical limits suggest
that $f_a\gtrsim0.6\times10^{9}$~GeV and
$m_a\lesssim0.01$~eV~\cite{Raffelt:1999tx,Eidelman:2004wy}.

Perhaps the most robust limits are those on the axion interaction
strength with photons and electrons because the properties of ordinary
stars can be used to test  anomalous energy losses caused by
processes such as $\gamma+e^-\to e^-+a$ or the Primakoff process
$\gamma+Ze\to Ze+a$ \cite{Raffelt:1999tx,Eidelman:2004wy}.  These
limits are so restrictive that processes involving the electron and
photon couplings will not be significant for the thermalization of
axions in the early universe in the post-QCD epoch.  On the other
hand, these interactions are also the most model dependent.
Axion-models can be constructed where the axion-photon interaction is
arbitrarily small.  In ``hadronic axion models'' such as the KSVZ
model~\cite{Kim:1979if,Shifman:1979if} there is no tree-level
interaction with ordinary quarks and leptons, and in any case, the
axion-electron coupling can be very small even in non-hadronic models
such as the DFSZ model~\cite{Zhitnitsky:1980tq,Dine:1981rt}.
Therefore, the restrictive stellar limits on the electron and photon
couplings do not conclusively rule out axions with a PQ scale in the
$10^6$~GeV range that is of interest to us.

The axion-nucleon interaction strength is constrained by two different
limits based on the observed neutrino signal of supernova 1987A
\cite{Raffelt:1999tx,Eidelman:2004wy}.  The axionic energy loss caused
by processes such as $NN\to NNa$ excludes a window of the
axion-nucleon interaction strength where the axionic contribution to
the loss or transfer of energy would have been comparable to or larger
than that of neutrinos. For a sufficiently large interaction strength,
axions no longer compete with neutrinos for the overall energy
transfer in the supernova core, but then would cause too many events
in the water Cherenkov detectors that observed the neutrino signal.
However, there is a narrow intermediate range of couplings,
corresponding to $f_a$ around $10^6$~GeV, i.e.\ to an axion mass of a
few~eV, where neither argument is conclusive.  In this ``hadronic
axion window,'' these elusive particles may still be allowed. This
observation has previously led to the speculation that axions with
eV-masses could contribute a significant hot dark-matter fraction to
the universe~\cite{Moroi:1998qs}.

\clearpage

Assuming that axions do not couple to charged leptons, the main
thermalization processes in the post-QCD epoch involve
hadrons~\cite{Chang:1993gm}
\begin{eqnarray}\label{eq:processes}
a+\pi\leftrightarrow\pi+\pi\,,\label{eq:pionprocess}\\
a+N\leftrightarrow N+\pi\,.
\end{eqnarray}
Due to the scarcity of nucleons relative to pions, the pionic
processes are actually by far the most important.  Therefore, we need
the axion-pion interaction which is given by a Lagrangian of the
form~\cite{Chang:1993gm}
\begin{equation}\label{eq:axionpionlagrangian}
{\cal L}_{a\pi}=\frac{C_{a\pi}}{f_\pi f_a}\,
\left(\pi^0\pi^+\partial_\mu\pi^-
+\pi^0\pi^-\partial_\mu\pi^+
-2\pi^+\pi^-\partial_\mu\pi^0\right)
\partial_\mu a\,.
\end{equation}
In hadronic axion models where the ordinary quarks and leptons do not
carry PQ charges, the coupling constant is~\cite{Chang:1993gm}
\begin{equation}\label{eq:axionpioncoupling}
C_{a\pi}=\frac{1-z}{3\,(1+z)}\,.
\end{equation}
In non-hadronic models an additional term enters that could, in
principle, reduce $C_{a\pi}$.  For the DFSZ model, the relevant
$a$-$\pi^0$-mixing strength was derived by Carena and
Peccei~\cite{Carena:1988kr}. In the DFSZ model, however, axions in our
$f_a$ range are already excluded so that we mainly focus on hadronic
models.

\section{Axion decoupling}                      \label{sec:freeze-out}

\subsection{Pionic process}

Axions are an often-cited cold dark matter candidate if they interact
so weakly (if $f_a$ is so large) that they never achieve thermal
equilibrium~\cite{Preskill:1982cy,Abbott:1982af,Dine:1982ah,%
  Davis:1986xc,Bradley:2003kg}. In that case coherent oscillations of
the axion field are excited around the QCD transition when the PQ
symmetry is explicitly broken by instanton effects. However, if $f_a$
is sufficiently small so that axions interact sufficiently strongly,
they will thermalize and survive as thermal relics in analogy to
neutrinos~\cite{Chang:1993gm,Turner:1986tb,Masso:2002np}. In the
eV-range of axion masses that is of interest to us, thermalization
will happen after the QCD transition so that one has to consider axion
interactions with hadrons rather than interactions with quarks and
gluons that would be relevant at earlier epochs.

Axions freeze out when their rate of interaction becomes slow compared
to the cosmic expansion rate.  As a criterion for axion thermal
decoupling we use
\begin{equation}\label{eq:freezeoutcondition}
\langle \Gamma_a\rangle_T=H(T)\,,
\end{equation}
where $\langle\Gamma_a\rangle_T$ is the axion absorption rate,
averaged over a thermal distribution at temperature $T$, whereas
$H(T)$ is the Hubble expansion parameter at the cosmic temperature
$T$. Our freeze-out criterion is accurate up to a constant of order
unity. In principle, the full system of Boltzmann equations should be
followed for all species, but the use of $H(T) =
\langle\Gamma_a\rangle_T$ only introduces an error of 10--20\% in the
decoupling temperature, consistent with the other approximations we
will use.

As explained in Sec.~\ref{sec:axions}, by far the most important
process for axion thermalization is the pionic reaction of
Eq.~(\ref{eq:pionprocess}) which is of the form $1+2\to 3+4$ with
particle 1 the axion and particles 2--4 the pions.  The average
absorption rate is
\begin{eqnarray}\label{eq:pionabsorptionrate1}
\langle\Gamma_a\rangle_T&=&\frac{1}{n_a}
\int\frac{d^3{\bf p}_1}{(2\pi)^32E_1}\,
\frac{d^3{\bf p}_2}{(2\pi)^32E_2}\,
\frac{d^3{\bf p}_3}{(2\pi)^32E_3}\,
\frac{d^3{\bf p}_4}{(2\pi)^32E_4}\,\sum|{\cal M}|^2\nonumber\\
&&\hskip2em{}\times(2\pi)^4\,
\delta^4(p_1+p_2-p_3-p_4)\, f_1f_2(1+f_3)(1+f_4)\,,
\end{eqnarray}
where $f_j$ are the thermal occupation numbers. The axion number
density in thermal equilibrium is $n_a=\int d^3{\bf p}_1
f_1/(2\pi)^3= (\zeta_3/\pi^2)\,T^3$.

The interaction Eq.~(\ref{eq:axionpionlagrangian}) implies the three
pionic processes $a+\pi^0\to \pi^++\pi^-$, $a+\pi^+\to \pi^++\pi^0$,
and $a+\pi^-\to \pi^-+\pi^0$. Assuming equal mass $m_\pi$ for both
charged and neutral pions, we find for the squared matrix element,
summed over all three channels,
\begin{equation}
\sum|{\cal M}|^2=\frac{9}{4}\,\left(\frac{C_{a\pi}}{f_af_\pi}\right)^2
\left(s^2+t^2+u^2-3m_\pi^4\right)\,,
\end{equation}
where $s=(p_1+p_2)^2$, $t=(p_1-p_3)^2$, and $u=(p_1-p_4)^2$.

\begin{figure}[ht]
\begin{indented}
\item[]
\includegraphics[width=100mm]{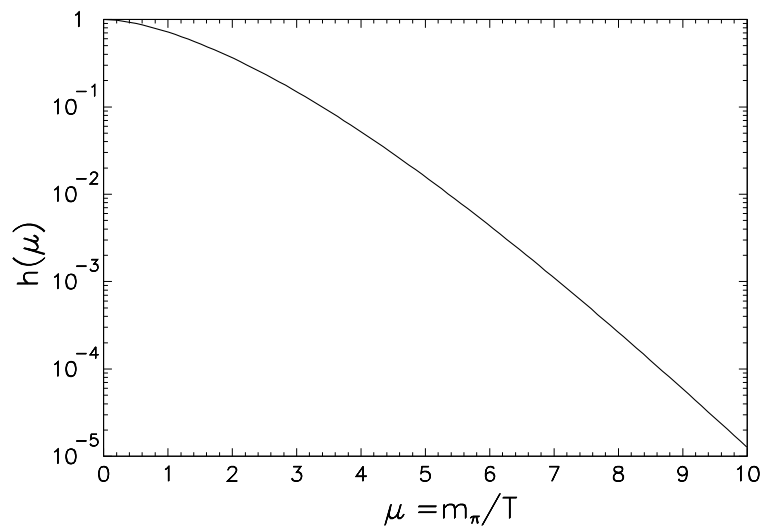}
\end{indented}
\caption{Dimensionless axion absorption rate $h(\mu)$ as defined
by Eq.~(\ref{eq:hdef}).}
\label{fig:hpion}
\end{figure}

We have performed the phase-space integration in
Eq.~(\ref{eq:pionabsorptionrate1}) along the lines of the method
described in Ref.~\cite{Hannestad:1995rs}. Chang and
Choi~\cite{Chang:1993gm} give an explicit three-dimensional integral
expression for the average rate of $\pi\pi\to\pi a$ when the
final-state stimulation factors $(1+f_3)(1+f_4)$ are neglected. Our
numerical integration agrees with their result if we consider the same
quantity, but in the following we use our full expression for the
average absorption rate.  From dimensional considerations one finds
\begin{equation}\label{eq:hdef}
\langle \Gamma_a\rangle_T=
A\,\left(\frac{C_{a\pi}}{f_af_\pi}\right)^2\,T^5\,h(m_\pi/T)\,,
\end{equation}
where numerically $A=0.215$.  The function $h(\mu)$
is normalized to $h(0)=1$ (Fig.~\ref{fig:hpion}).

\subsection{Expansion rate}

The cosmic expansion rate is given by the Friedmann equation as
$H^2=8\pi G_{\rm N}\rho/3$ with $G_{\rm N}$ the Newton constant and
$\rho$ the mass-energy density. For the radiation dominated epoch one
usually writes~\cite{Kolb:1990vq}
\begin{equation}\label{eq:energydensity}
\rho=\frac{\pi^2}{30}\,g_*(T)\,T^4
=\sum_{j={\rm particles}}g_j
\int\frac{d^3{\bf p}}{(2\pi)^3}
\frac{(m_j^2+{\bf p}^2)^{1/2}}
{\exp\left[(m_j^2+{\bf p}^2)^{1/2}/T\right]\pm1}
\end{equation}
where $g_*(T)$ denotes the effective number of thermal degrees of
freedom that are excited at the epoch with temperature $T$.  Further,
$g_j$ is the number of internal degrees of freedom of a given particle
species $j$ with mass $m_j$. The $\pm1$ in the denominator applies to
bosons and fermions, respectively. We assume vanishing chemical
potentials for all particles. For the conditions of interest, the
particle-antiparticle asymmetry is small even for nucleons.  The
Friedmann equation is thus
\begin{equation}
H=\left[\frac{4\pi^3}{45}\,g_*(T)\right]^{1/2}\,
\frac{T^2}{m_{\rm Pl}}\,,
\end{equation}
where $m_{\rm Pl}=G_{\rm N}^{-1/2}$ is the Planck mass.

\begin{figure}[ht]
\begin{indented}
\item[]
\includegraphics[width=100mm]{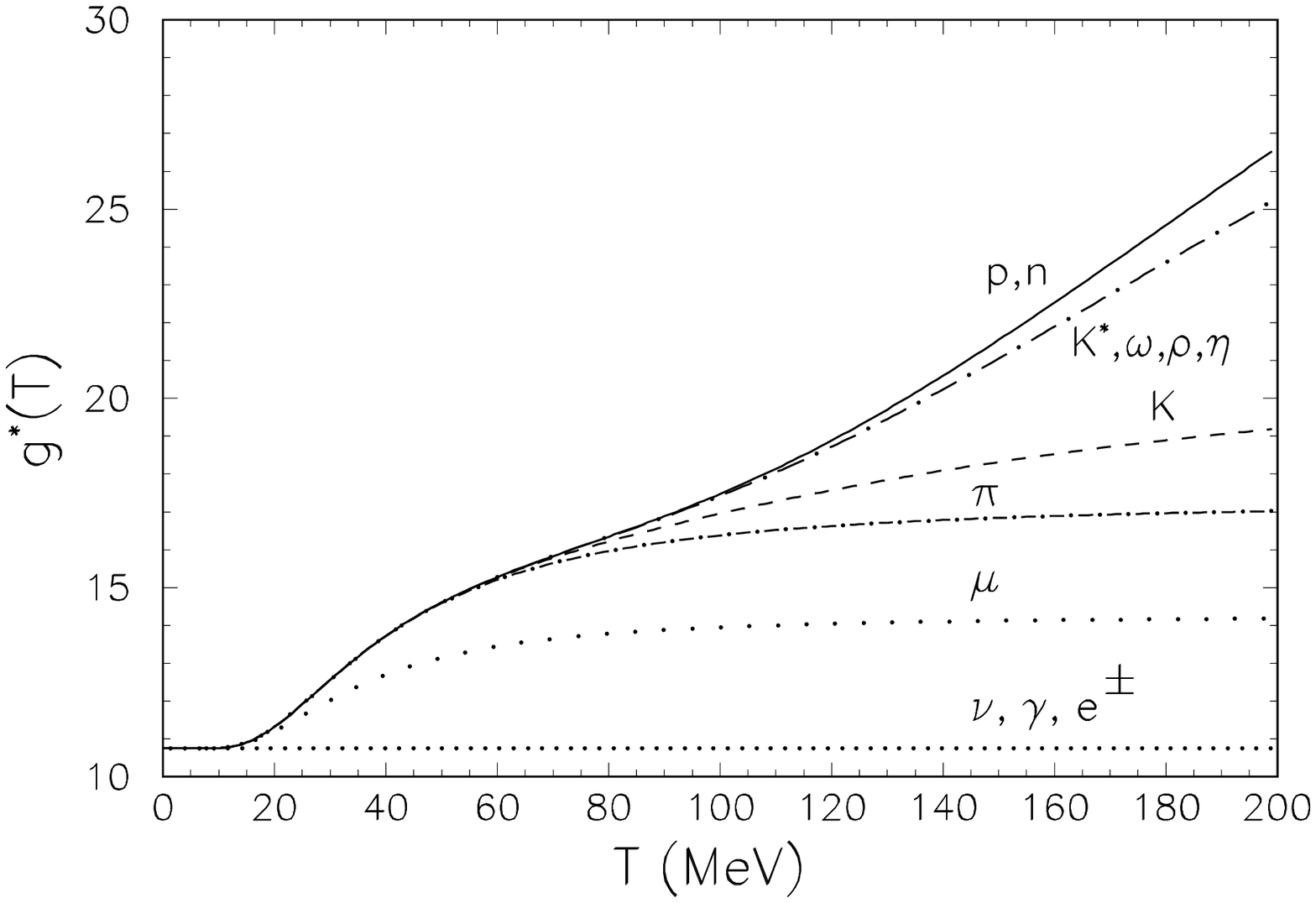}
\item[]
\includegraphics[width=100mm]{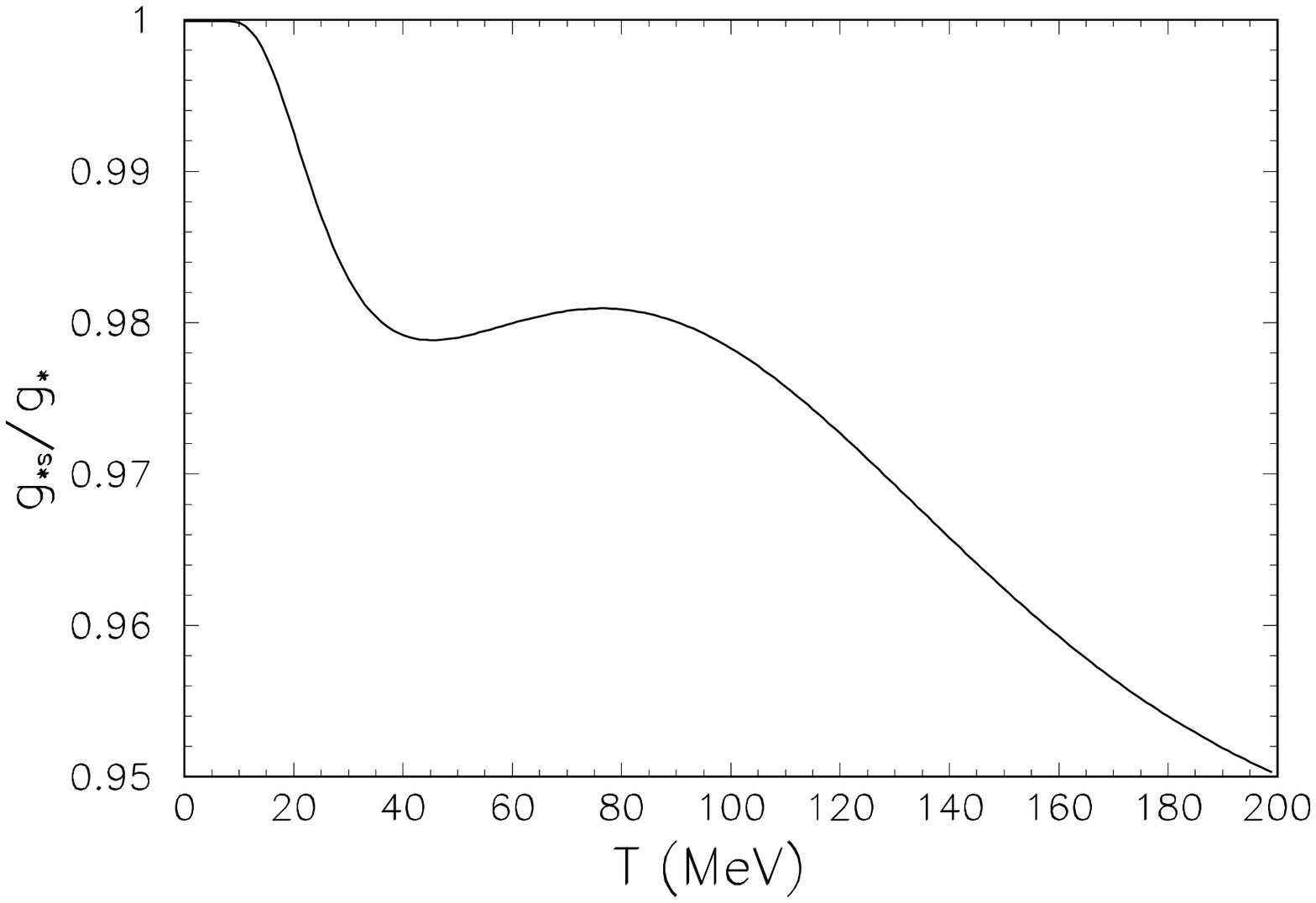}
\end{indented}
\caption{Effective number of thermal degrees of
freedom in the early universe during the post-QCD epoch, assuming the
particle content listed in Table~\ref{tab:particles}.
{\it Upper panel:} $g_*$. {\it Lower Panel:} $g_{*S}/g_*$.}
\label{fig:thermaldegrees}
\end{figure}

We consider all particles with masses up to the nucleon mass as listed
in Table~\ref{tab:particles}. In the
upper panel of Fig.~\ref{fig:thermaldegrees} we show the function
$g_*(T)$, assuming that only the particles of
Table~\ref{tab:particles} contribute and ignoring the color
deconfinement transition that probably takes place at a temperature
somewhat below 200~MeV.

\begin{table}[ht]
\caption{\label{tab:particles}Particles that contribute to $g_*$
  during the post-QCD epoch}
\begin{indented}
\item[]
\begin{tabular}{@{}llll}
\br
Particles&Mass [MeV]&Spin&$g_j$\\
\mr
$\gamma$&0&$1$&2\\
$\nu$, $\bar\nu$&0&$\frac{1}{2}$&6\\
$e^\pm$&0.511&$\frac{1}{2}$&4\\
$\mu^\pm$&106&$\frac{1}{2}$&4\\
$\pi^0$, $\pi^\pm$&138&$0$&3\\
$K^0$, $\bar K^0$, $K^\pm$&494&$0$&4\\
$\eta$&547&$0$&1\\
$\rho^0$, $\rho^\pm$&771&1&9\\
$\omega$&782&$1$&3\\
$K^{*0}$, $\bar K^{*0}$, $K^{*\pm}$&892&$1$&12\\
$n$, $p$, $\bar n$, $\bar p$&938&$\frac{1}{2}$&8\\
\br
\end{tabular}
\end{indented}
\end{table}

The effective number of thermal degrees of freedom $g_*$ that appears
in the Friedmann equation is the quantity that governs the expansion
rate. However, if axions freeze out at a certain epoch, their number
density at late times is governed by $g_{*S}$, the effective number of
thermal degrees characterizing the entropy at the freeze-out epoch.
The entropy density of an ideal gas is given by the general expression
$(\rho+P)/T\propto g_{*S} T^3$ with $P$ the pressure.  The pressure is
given by the same integral expression as the energy density in
Eq.~(\ref{eq:energydensity}) after replacing $E_j=(m_j^2+{\bf
  p}^2)^{1/2}$ by $\frac{1}{3}\,{\bf p}\cdot{\bf v}= \frac{1}{3}{\bf
  p}^2/E_j$. Even though we assume that all particles are in thermal
equilibrium at the same temperature, there will be a difference
between $g_*$ and $g_{*S}$ because some of the contributing particles
are not massless. In the lower panel of Fig.~\ref{fig:thermaldegrees}
we show the ratio $g_{*S}/g_*$.
Since the deviation of $g_*$ from $g_{*S}$ is at most a few percent
for the conditions of interest, we will henceforth ignore the
difference between the two quantities and always use $g_*$.  Moreover,
since axions themselves contribute only a single degree of freedom we
neglect their contribution to $g_*$ for simplicity.

\subsection{Freeze-out conditions}

\begin{figure}[ht]
\begin{indented}
\item[]
\includegraphics[width=100mm]{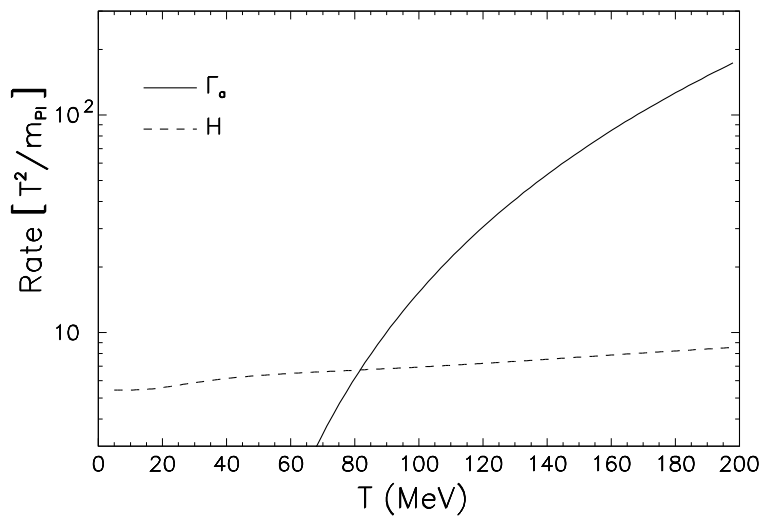}
\end{indented}
\caption{Average axion absorption rate for $f_a=10^7$~GeV and cosmic
  expansion rate as a function of the cosmic temperature. Rates
  are in units of $T^2/m_{\rm Pl}$.}
\label{fig:GammaH}
\end{figure}

\begin{figure}[ht]
\begin{indented}
\item[]
\includegraphics[width=100mm]{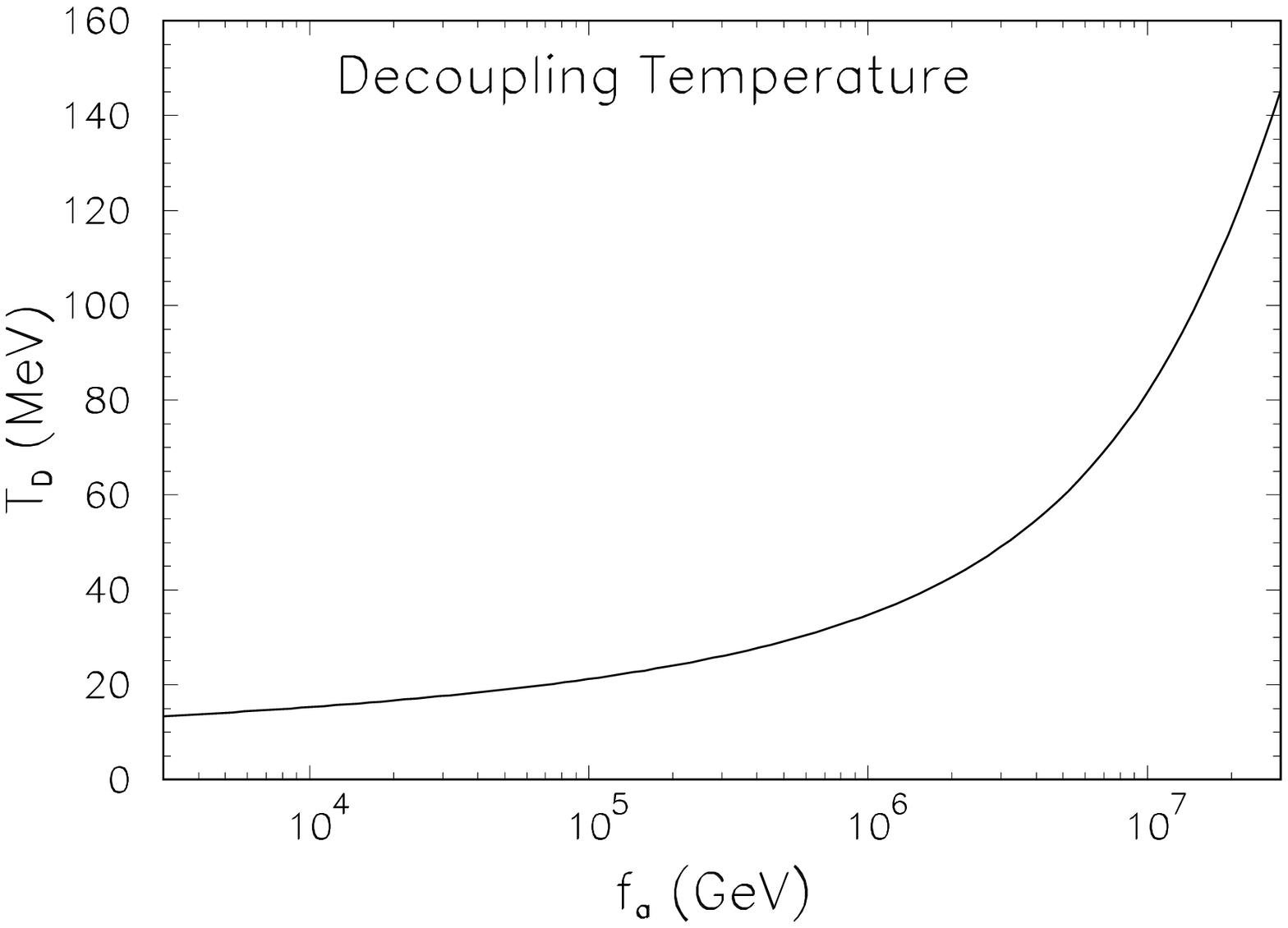}
\item[]
\includegraphics[width=100mm]{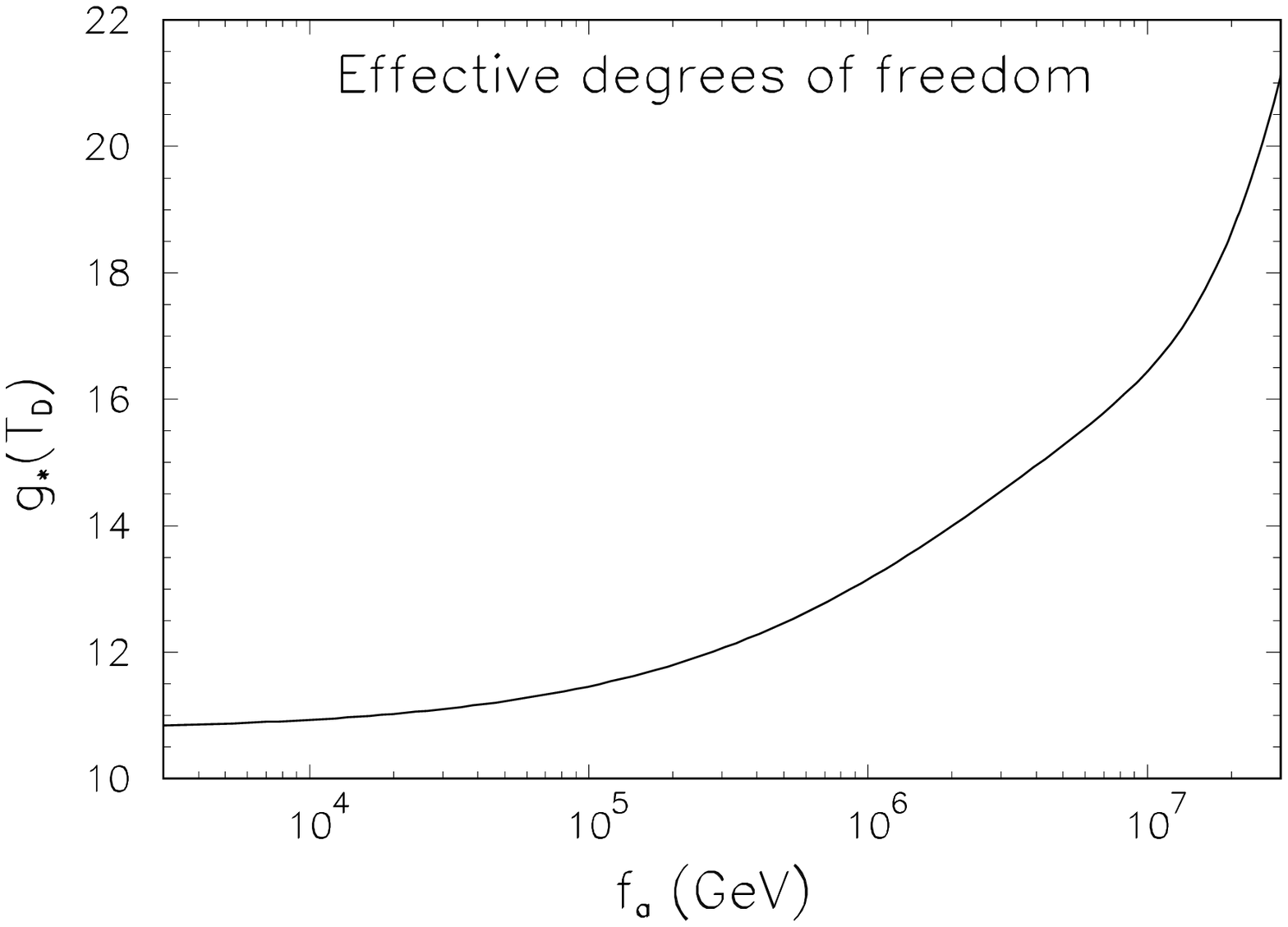}
\item[]
\includegraphics[width=100mm]{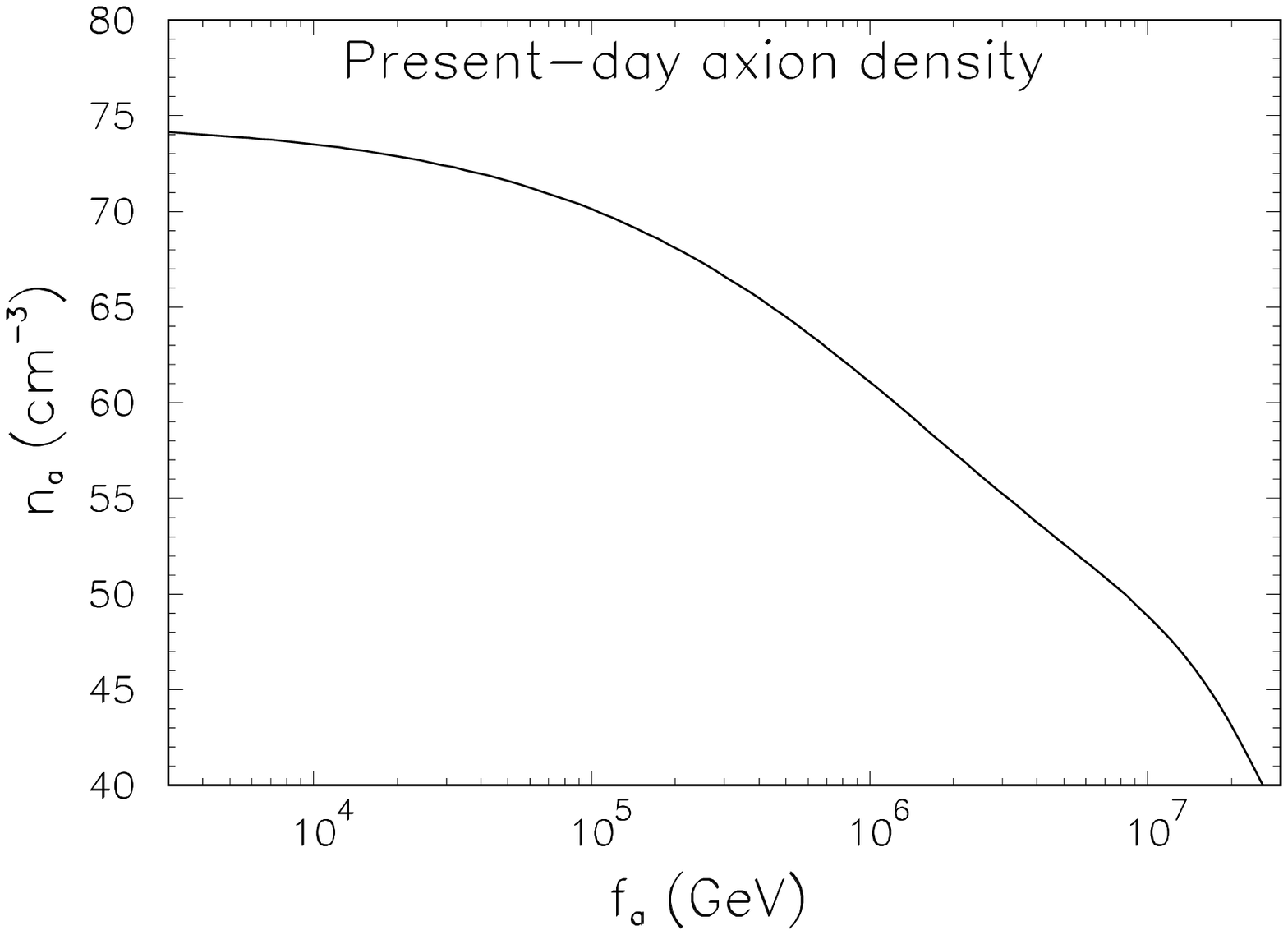}
\end{indented}
\caption{Axion freeze-out, assuming the
$a$-$\pi$ coupling of Eq.~(\ref{eq:axionpioncoupling}).
{\it Top:} Decoupling temperature.
{\it Middle:} Corresponding $g_*$.
{\it Bottom:} Present-day axion density.}
\label{fig:decoupling}
\end{figure}

We now combine our result for the cosmic expansion rate in the
post-QCD epoch with that for the axion absorption rate and determine
the freeze-out conditions from Eq.~(\ref{eq:freezeoutcondition}).  As
an example we show $H(T)$ and $\langle\Gamma_a\rangle_T$ in
Fig.~\ref{fig:GammaH}, assuming a PQ scale of $f_a=10^7$~GeV and the
hadronic axion-pion coupling of Eq.~(\ref{eq:axionpioncoupling}).
From the intersection point we determine $T_{\rm D}$, the axion
decoupling temperature, and the corresponding $g_*$. We show these
quantities as functions of $f_a$ in Fig.~\ref{fig:decoupling} and in
Table~\ref{tab:decoupling}.

Moreover, from $g_*$ at decoupling one can determine the present-day
number density of axions by virtue of the relation
\begin{equation}
n_a=\frac{g_{*S}({\rm today})}{g_{*S}({\rm decoupling})}\times
\,\frac{n_\gamma}{2}\,,
\end{equation}
where $n_\gamma=411~{\rm cm}^{-3}$~\cite{Eidelman:2004wy} is the
present-day density of cosmic microwave photons and today $g_{*S}=
3.91$ \cite{Kolb:1990vq}. We show $n_a$ as a function of $f_a$ in
Fig.~\ref{fig:decoupling} and give some numerical values in
Table~\ref{tab:decoupling}. For comparison, we note that the
present-day neutrino number density, determined by standard neutrino
freeze-out, is $n_\nu \simeq 112~{\rm cm}^{-3}$.

\begin{table}[ht]
\caption{\label{tab:decoupling}Thermal decoupling conditions for
axions, assuming the hadronic axion-pion coupling of
Eq.~(\ref{eq:axionpioncoupling}). $n_a$ is the expected present-day
axion density.}
\begin{indented}
\item[]
\begin{tabular}{@{}llll}
\br
$f_a$ [GeV]&$T_{\rm D}$ [MeV]&$g_*(T_{\rm D})$&$n_a$ [cm$^{-3}$]\\
\mr
$3\times10^3$& 13.37 & 10.84 & 74.14  \\
$1\times10^4$& 15.30 & 10.93 & 73.50  \\
$3\times10^4$&  17.63 & 11.10 & 72.39 \\
$1\times10^5$&  21.21 & 11.46 & 70.11 \\
$3\times10^5$&  26.06 & 12.06 & 66.63 \\
$1\times10^6$&  34.75 & 13.15 & 61.08 \\
$3\times10^6$&  49.12 & 14.54 & 55.24 \\
$1\times10^7$&  81.61 & 16.43 & 48.88 \\
$3\times10^7$&  145.31 & 21.10 & 38.08\\
\br
\end{tabular}
\end{indented}
\end{table}

\clearpage

\section{Likelihood analysis}                   \label{sec:likelihood}

\subsection{Theoretical predictions}

In order to derive limits on the axion parameters we compare
theoretical power spectra for the matter distribution and the CMB
temperature fluctuations with observational data in analogy to
previous works by one of
us~\cite{Hannestad:2002xv,Hannestad:2003xv,Hannestad:2004bu}.  The
predicted power spectra are calculated with the publicly available
CMBFAST package~\cite{CMBFAST}.  As in our previous paper on general
low-mass thermal relics~\cite{Hannestad:2003ye} we have modified the
code to include a scalar boson that decouples earlier than neutrinos,
i.e.\ we assume a number density and velocity distribution according
to the freeze-out epoch.

As a set of cosmological parameters apart from the axion
characteristics we choose the matter density $\Omega_{\rm M}$, the
baryon density $\Omega_{\rm B}$, the Hubble parameter $H_0$, the
scalar spectral index of the primordial fluctuation spectrum $n_s$,
the optical depth to reionization $\tau$, the normalization of the CMB
power spectrum $Q$, and the bias parameter~$b$.  We restrict our
analysis to geometrically flat models $\Omega = \Omega_{\rm M} +
\Omega_\Lambda = 1$. The cosmological parameters and their assumed
priors are listed in Table~\ref{tab:priors}.  The actual
marginalization over parameters was performed using a simulated
annealing procedure~\cite{Hannestad:wx}. The cosmological parameters
in our model corresponds to the simplest $\Lambda$CDM model which fits
present data. While our results would be changed slightly by including
additional parameters, there would be no significant
changes~\cite{Hannestad:2003ye}.

Likelihoods are calculated from $\chi^2$ so that for 1 parameter
estimates, 68\% confidence regions are determined by $\Delta \chi^2 =
\chi^2 - \chi_0^2 = 1$, and 95\% regions by $\Delta \chi^2 = 4$.
Here, $\chi_0^2$ refers to the best-fit model.

\begin{table}[ht]
\caption{\label{tab:priors} Priors on cosmological parameters used in
the likelihood analysis.}
\begin{indented}
\item[]
\begin{tabular}{@{}lll}
\br
Parameter &Prior&Distribution\cr
\mr
$\Omega=\Omega_{\rm M}+\Omega_\Lambda$&1&Fixed\\
$\Omega_{\rm M}$ & 0.1--1 & Top Hat\\
$h$ & $0.72 \pm 0.08$&Gaussian\\
$\Omega_{\rm B}h^2$ & 0.014--0.040&Top hat\\
$n_s$ & 0.6--1.4& Top hat\\
$\tau$ & 0--1 &Top hat\\
$Q$ & --- &Free\\
$b$ & --- &Free\\
\br
\end{tabular}
\end{indented}
\end{table}

\subsection{Cosmological Data}

\subsubsection{Large Scale Structure.}

At present there are two large galaxy surveys of comparable size, the
Sloan Digital Sky Survey (SDSS) \cite{Tegmark:2003uf,Tegmark:2003ud}
and the 2~degree Field Galaxy Redshift Survey (2dFGRS) \cite{2dFGRS}.
Once the SDSS is completed it will be significantly larger and more
accurate than the 2dFGRS. We will only use data from SDSS, but the
results would be almost identical with 2dFGRS data.  We use only data
on scales larger than $k = 0.15\,h~{\rm Mpc}^{-1}$ to avoid problems
with non-linearity.

\subsubsection{Cosmic Microwave Background.}

The CMB temperature fluctuations are conveniently described in terms
of the spherical harmonics power spectrum $C_l^{TT} \equiv \langle
|a_{lm}|^2 \rangle$, where $\frac{\Delta T}{T} (\theta,\phi) =
\sum_{lm} a_{lm}Y_{lm}(\theta,\phi)$.  Since Thomson scattering
polarizes light, there are also power spectra coming from the
polarization. The polarization can be divided into a curl-free $(E)$
and a curl $(B)$ component, yielding four independent power spectra:
$C_l^{TT}$, $C_l^{EE}$, $C_l^{BB}$, and the $T$-$E$ cross-correlation
$C_l^{TE}$.

The WMAP experiment has reported data only on $C_l^{TT}$ and
$C_l^{TE}$ as described in Refs.~\cite{Bennett:2003bz,Spergel:2003cb,%
Verde:2003ey,Kogut:2003et,Hinshaw:2003ex}.  We have performed our
likelihood analysis using the prescription given by the WMAP
collaboration~\cite{Spergel:2003cb,%
Verde:2003ey,Kogut:2003et,Hinshaw:2003ex} which includes the
correlation between different $C_l$'s. Foreground contamination has
already been subtracted from their published data.

\subsubsection{Supernova luminosity distances.}

We use the ``gold'' data set compiled and described by Riess {\it et
al.}~\cite{Riess:2004} consisting of 157 SNe~Ia using a modified
version of the SNOC package~\cite{Goobar:2002c}.

\subsubsection{The Lyman-$\alpha$ forest.}

We have furthermore used the Lyman-$\alpha$ forest data from Croft
{\it et al.} \cite{lya}. The error bars on the last three data points
have been increased in the same fashion as was done by the WMAP
collaboration~\cite{Bennett:2003bz,Spergel:2003cb,Verde:2003ey,%
Kogut:2003et,Hinshaw:2003ex}, in order to make them compatible with
the analysis of Gnedin and Hamilton~\cite{gneha}.  While there is some
controversy about the use of Lyman-$\alpha$ data for parameter
estimation it should be noted that their inclusion has only a very
small quantitative effect on our results as will become clear below.

\subsubsection{Hubble parameter.}

Additionally we use data on the Hubble parameter from the HST key
project \cite{Freedman:2000cf}, $h = H_0/(100~{\rm km}~{\rm
s}^{-1}~{\rm Mpc}^{-1}) = 0.72 \pm 0.08$.

\subsection{Limits on axion parameters}

\begin{figure}[ht]
\begin{indented}
\item[]
\includegraphics[width=100mm]{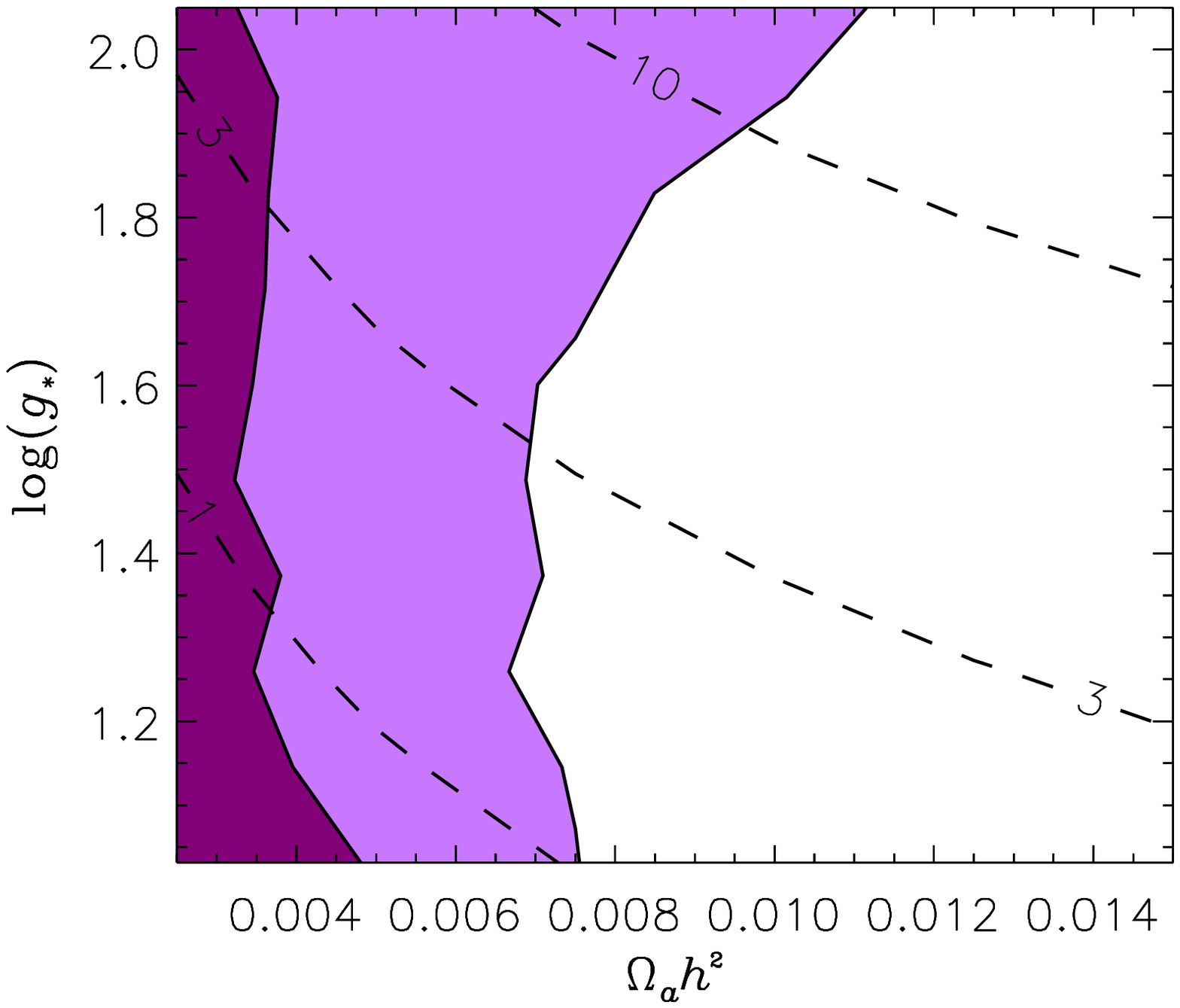}
\item[]
\includegraphics[width=100mm]{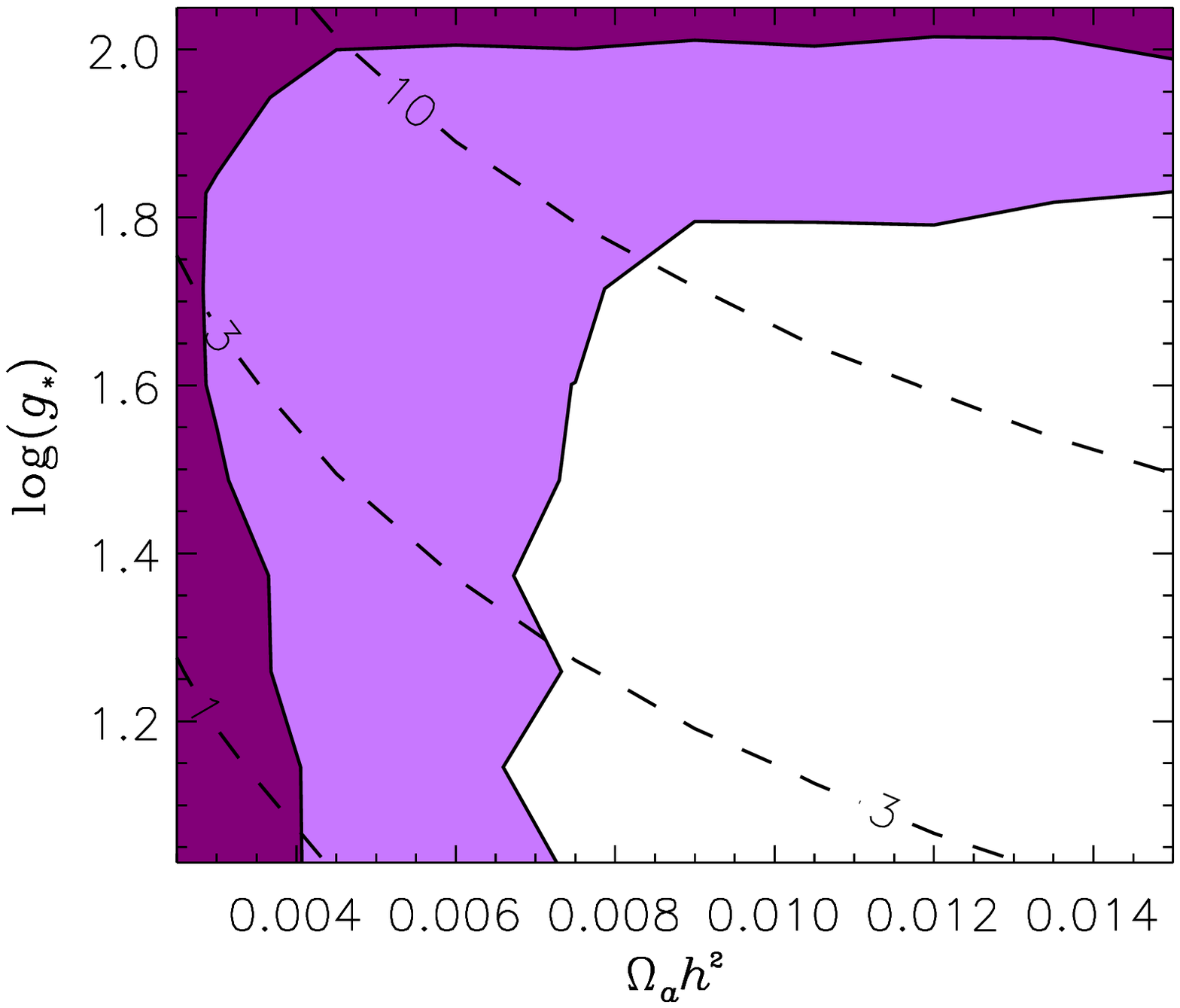}
\end{indented}
\caption{Likelihood contours for the allowed axion
  parameters. Everything to the right of the dark shaded region is
  excluded at the 68\% CL, and everything to the right of the light
  shaded region is excluded at the 95\% CL. Contours of equal axion
  mass with the indicated $m_a$ values in~eV are shown as dashed
  lines.  {\it Upper panel:}\/ All data included.  {\it Lower
    panel:}\/ Lyman-$\alpha$ data excluded.}
\label{fig:axion1}
\end{figure}

After marginalizing over the cosmological parameters shown in
Table~\ref{tab:priors}, the 68\% and 95\% CL allowed regions of axion
parameters are shown in Fig.~\ref{fig:axion1}. In the upper panel we
have included the full data set described above, while in the lower
panel we have removed the Lyman-$\alpha$ data that is perhaps our most
uncertain input.

The Lyman-$\alpha$ data only has a significant impact for high values
of $g_*$ because it measures very small scales. The large-scale
structure data loses sensitivity at $k \lesssim 0.15\,h~{\rm
Mpc}^{-1}$, whereas the Lyman-$\alpha$ data probes scales which are
about an order of magnitude smaller. For high $g_*$, the particle mass
for a given value of $\Omega_a h^2$ is higher and therefore the
free-streaming length is correspondingly smaller. The change in power
spectrum amplitude occurs at $k > 0.15\,h~{\rm Mpc}^{-1}$ and is only
measurable in the Lyman-$\alpha$ data when $g_* \gtrsim 60$--70.

\begin{figure}[ht]
\begin{indented}
\item[]
\includegraphics[width=100mm]{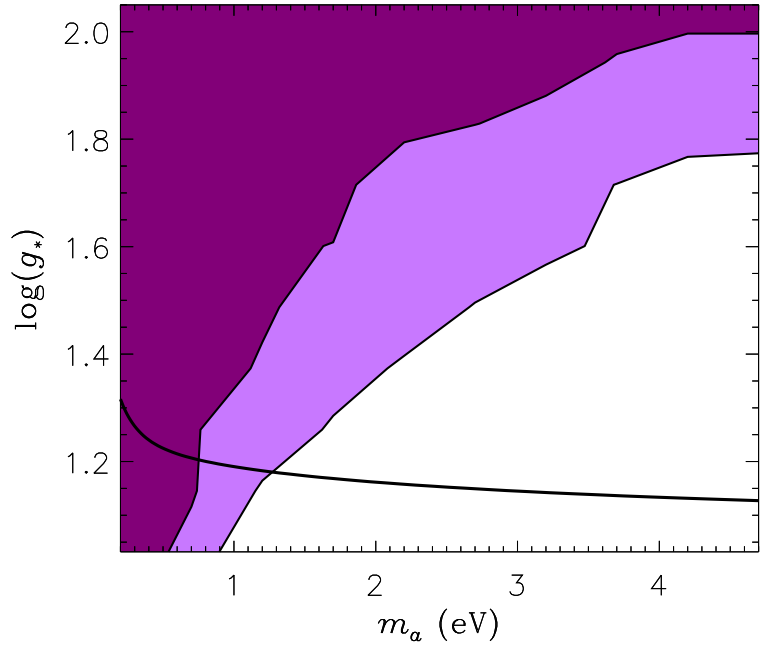}
\end{indented}
\caption{Likelihood contours including all data, equivalent
  to the upper panel of Fig.~\ref{fig:axion1}, transformed to the
  parameter plane of $g_*$ and $m_a$.
  The relation between $g_*$ and $m_a$ for hadronic axions
  is shown as a thick solid line.}
\label{fig:axion2}
\end{figure}

In Fig.~\ref{fig:axion2} we show the same analysis as in the upper
panel of Fig.~\ref{fig:axion1}, but now transformed to the
$m_a$-$g_*$-plane.  The hadronic axion model is shown as a thick solid
line. It lies at relatively small values of $g_*$ for the relevant
mass range, meaning that there will be very little quantitative
difference if the Lyman-$\alpha$ data is not used.

\begin{figure}[ht]
\begin{indented}
\item[]
\includegraphics[width=100mm]{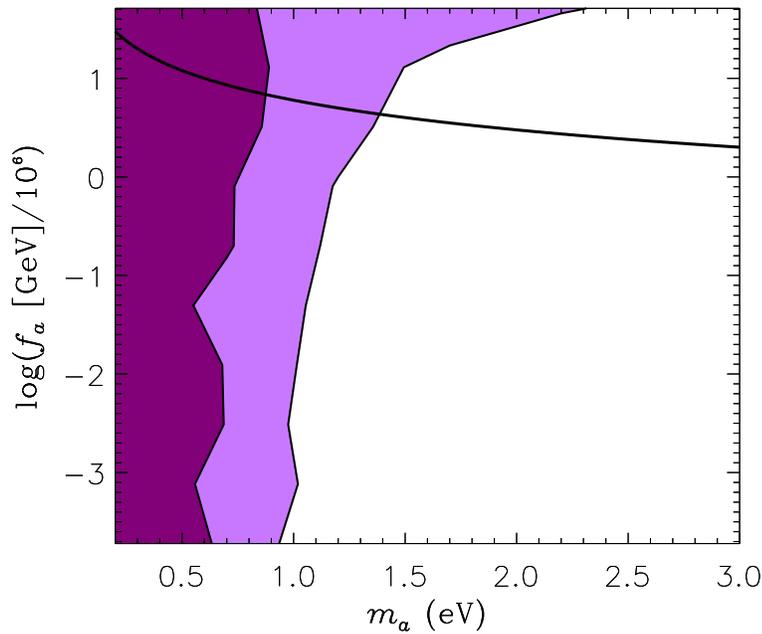}
\end{indented}
\caption{Likelihood contours including all data, equivalent to the
  upper panel of Fig.~\ref{fig:axion1}, transformed to the parameter
  plane of $f_a$ and $m_a$. We have used the connection between $g_*$
  and $f_a$ for hadronic axions as shown in Fig.~\ref{fig:decoupling}
  and Table~\ref{tab:decoupling}.  The standard $m_a$-$f_a$-relation
  of Eq.~(\ref{eq:axmass}) is shown as a thick solid line.}
\label{fig:axion3}
\end{figure}

\clearpage

Likewise, in Fig.~\ref{fig:axion3} we show the analysis in the
$m_a$-$f_a$-plane, where the relationship between $f_a$ and $g_*$
shown in Fig.~\ref{fig:decoupling} and Table~\ref{tab:decoupling} has
been used. In this figure, the true independent variable on the
vertical axis is the axion-pion coupling that was transformed to a
value for the PQ scale assuming the hadronic case
Eq.~(\ref{eq:axionpioncoupling}). For other models with different
$C_{a\pi}$, the vertical axis in Fig.~\ref{fig:axion3} must be
re-scaled accordingly. Moreover, the standard relationship between
$m_a$ and $f_a$ of Eq.~(\ref{eq:axmass}) and the hadronic axion-pion
coupling of Eq.~(\ref{eq:axionpioncoupling}) put axions on the thick
solid line.

\begin{figure}[ht]
\begin{indented}
\item[]
\includegraphics[width=100mm]{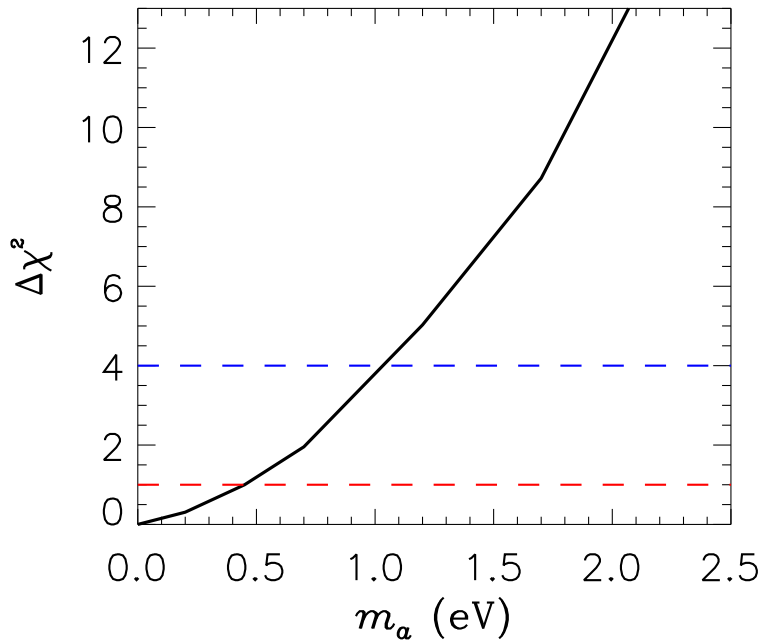}
\end{indented}
\caption{One-dimensional likelihood analysis for the hadronic axion
  model obeying Eq.~(\ref{eq:axmass}). Shown is $\Delta \chi^2 =
  \chi^2 - \chi_0^2$ as a function of $m_a$ where $\chi_0^2$ refers to
  the best-fit model. In this 1D analysis the 95\% excluded region is
  at $\Delta \chi^2 > 4$, and the 68\% excluded region at $\Delta
  \chi^2 > 1$, as indicated by the horizontal dashed lines.}
\label{fig:axion4}
\end{figure}

In this case, where the most specific assumptions about the underlying
axion model were made, the axion parameter space collapses to one
dimension, leading to a more restrictive mass limit.  In
Fig.~\ref{fig:axion4} we show a one-dimensional likelihood analysis
for this case, i.e.\ for the hadronic axion model.  As a result one
finds a 95\% CL excluded region for $m_a$ of
\begin{equation}
m_a < 1.05~{\rm eV} \quad (95\%~{\rm CL}),
\end{equation}
based on all the available observational data. For comparison we note
that the same data and the same analysis method provide
\begin{equation}
\sum m_\nu < 0.65~{\rm eV} \quad(95\%~{\rm CL})
\end{equation}
for neutrinos~\cite{Hannestad:2004bu}.

\section{Summary}                                  \label{sec:summary}

We have studied structure-formation limits on axions or other scalar
particles that have a small mass and that couple to pions by virtue of
the Lagrangian Eq.~(\ref{eq:axionpionlagrangian}).  Our main result is
the exclusion range in the two-dimensional $f_a$-$m_a$ parameter space
shown in Fig.~\ref{fig:axion3} where the axion-pion coupling strength
of Eq.~(\ref{eq:axionpioncoupling}) was assumed, i.e.\ this is an
exclusion range in the parameter space of axion mass and axion-pion
coupling. Assuming in addition the standard $m_a$-$f_a$ relationship
of Eq.~(\ref{eq:axmass}), we find a limit on the axion mass of
$m_a<1.05$~eV (95\% CL), corresponding to $f_a>5.7\times 10^6$~GeV.

A comparable axion mass limit is obtained from the requirement that
excessive energy losses of horizontal branch stars in globular
clusters should be avoided~\cite{Raffelt:1999tx}. However, the
globular-cluster limit depends on the axion-photon interaction that is
rather model dependent.

The neutrino observations of supernova 1987A provide restrictive
limits on the axion-nucleon interaction, suggesting
$m_a\lesssim0.01$~eV or $f_a\gtrsim0.6\times10^9$~GeV if one assumes
generic coupling strengths between axions and nucleons. However, the
supernova argument may leave open the ``hadronic axion window'' of
about $3\times10^{5}~{\rm GeV}\lesssim f_a\lesssim 3\times10^{6}~{\rm
GeV}$ \cite{Moroi:1998qs,Chang:1993gm}. Our result closes this window,
at least for generic values of the axion-nucleon and axion-pion
couplings.

More importantly, we have provided a new limit on axion parameters
based on a different interaction channel than previous limits and
based on different data and assumptions. Every experimental
measurement and every astrophysical or cosmological argument has its
own systematic uncertainties and its own recognized or un-recognized
loop holes. Therefore, to corner axions it is certainly important to
use as many independent interaction channels and as many different
approaches as possible.

An ongoing experimental axion search, the CERN Axion Solar Telescope
(CAST), is based on the axion-photon interaction.  It has reported
first limits for $m_a\lesssim0.02$~eV \cite{cast}. In the second phase
it will cover axion masses up to approximately 1~eV.  It is noteworthy
that our new limit is almost identical with the upper mass range that
can be reached with CAST, i.e.\ the CAST search and our new limit are
nicely complementary.

\section*{Acknowledgments} 

We acknowledge use of the publicly available CMBFAST
package~\cite{CMBFAST} and of computing resources at DCSC (Danish
Center for Scientific Computing).  A.M. thanks P.~Serpico and
S.~Uccirati for useful comments and fruitful discussions.  In Munich,
this work was supported, in part, by the Deutsche
Forschungsgemeinschaft (DFG) under grant No.~SFB-375.  The work of
A.M. is supported in part by the Italian ``Istituto Nazionale di
Fisica Nucleare'' (INFN) and by the ``Ministero dell'Istruzione,
Universit\`a e Ricerca'' (MIUR) through the ``Astroparticle Physics''
research project.

\clearpage

\section*{References} 

\end{document}